\documentclass[year=24,pdfa]{fmcad}

\usepackage[normalem]{ulem}

\usepackage{amsmath,amssymb,amsfonts}
\usepackage{pgffor}
\usepackage{xcolor}

\usepackage{newfloat}
\usepackage{syntax}
\usepackage{xpatch}
\usepackage{subcaption}
\usepackage{cite}
\usepackage{caption}    
\usepackage{listings}
\usepackage{xspace}
\usepackage{graphicx}
\usepackage{calc}
\usepackage{ragged2e}
\usepackage{scrextend}
\usepackage{changepage}
\usepackage{bm}
\usepackage{comment}
\usepackage{booktabs}
\usepackage{array}
\usepackage{mathtools}
\usepackage{multicol}
\usepackage{stmaryrd}
\usepackage{rotating}
\usepackage{multirow}
\usepackage{enumitem}
\usepackage{stackengine}
\usepackage{ebproof}
\usepackage{xspace}
\usepackage[noend]{algpseudocode}

\usepackage{hyperref}

\definecolor{codeblue}{rgb}{0.2,0.2,1}
\definecolor{codegreen}{rgb}{0,0.6,0}
\definecolor{codegray}{rgb}{0.5,0.5,0.5}
\definecolor{codepurple}{rgb}{0.58,0,0.82}
\definecolor{codeorange}{rgb}{.8,0.3,0.1}
\makeatletter
\lst@InstallKeywords k{assert}{assertstyle}{}{basicstyle}{}ld
\makeatother

\makeatletter
\lst@InstallKeywords k{adt}{adtstyle}{}{basicstyle}{}ld
\makeatother

\makeatletter
\def\lst@Literatekey#1\@nil@{\let\lst@ifxliterate\lst@if
    \expandafter\def\expandafter\lst@literate\expandafter{\lst@literate#1}}
\makeatother
\newcommand{\smaller}{\fontsize{7}{8.3}\selectfont}

\lstdefinestyle{pystyle}{
    language=Python,
    literate={\ \ \ \ }{{\ \ }}{2},
    commentstyle=\color{codegreen},
    keywordstyle=\color{codeblue}\bfseries,
    numberstyle=\tiny\color{codegray},
    stringstyle=\color{codepurple},
    basicstyle=\small\ttfamily,
    adtstyle=\color{codeorange}\bfseries,
    moreadt={Sum,Product,Enum},
    tabsize=2,
    deletekeywords=[2]{format,sum,abs,float},
    morekeywords={assert},
    numberstyle=\smaller, 
}

\lstdefinestyle{constraint}{
    style=pystyle,
    assertstyle=\color{codepurple}\bfseries,
    moreassert={EXISTS,FORALL,IF,ASSERT,LET},
    morekeywords={foreach}
}
    
\lstdefinestyle{inline}{
    style=constraint,
    basicstyle=\ttfamily,
    deletekeywords={and},
}

\DeclareCaptionLabelFormat{center}{\hspace*{\fill}#1 #2\hspace*{\fill}}

\DeclareCaptionStyle{capstyle}{
    style=base,
    textfont={normal,small},
    labelfont=bf,
    margin=.01\columnwidth,
}

\newcommand{\mton}{many-to-many\xspace} 
\newcommand{\genall}{\ensuremath{\mathit{genAll}}\xspace}
\newcommand{\genlc}{\ensuremath{\mathit{genAll}_{LC}}\xspace}
\newcommand{\IterativeCEGIS}{\ensuremath{\mathit{IterativeCEGIS}}\xspace}
\newcommand{\baseline}{\ensuremath{\mathit{IterativeCEGIS_{GCBPS}}}\xspace}

\newcommand{\vv}[1]{\mathbf{#1}\xspace}
\newcommand{\ir}[1]{{\vv{#1}}^{\mathit{IR}}\xspace}
\newcommand{\isa}[1]{{\vv{#1}}^{\mathit{ISA}}\xspace}
\newcommand{\vva}[1]{{\vv{#1}}^{a}\xspace}
\newcommand{\vvb}[1]{{\vv{#1}}^{b}\xspace}

\newcommand{\wfp}{\ensuremath{\mathit{wfp}}\xspace}
\newcommand{\prog}{\ensuremath{\mathit{prog}}\xspace}
\newcommand{\synth}{\ensuremath{\mathit{synth}}\xspace}
\newcommand{\verif}{\ensuremath{\mathit{verif}}\xspace}
\newcommand{\CEGIS}{\mathit{CEGIS}\xspace}

\newcommand{\ruleeq}{\sim_{rule}\xspace}

\newcommand{\ruleset}[1]{[#1]_{rule}}
\newcommand{\PatL}{\psi_{pat}\xspace} 
\newcommand{\RuleL}{\psi_{\mathit{dup}}\xspace}

\newcommand{\boldS}[1]{\mathbb{S}_{#1}}

\newcommand{\model}{\mathcal{M}\xspace}
\newcommand{\GCBPS}{\mathit{GCBPS}\xspace}
\newcommand{\CBPS}{CBPS\xspace}
\newcommand{\SR}{\boldS{R}\xspace}

\newcommand{\AllC}{\mathit{AllComposites}\xspace}
\newcommand{\multicomb}{\mathit{multicomb}\xspace}
\newcommand{\allInputs}{\mathit{allInputs}\xspace}
\newcommand{\CEGISAll}{\mathit{CEGISAll}\xspace}
\newcommand{\constructRule}{\mathit{rewriteRule}\xspace}
\newcommand{\cost}{\mathit{cost}\xspace}
\newcommand{\sorted}{\mathit{sorted}\xspace}
\newcommand{\sortByCost}{\mathit{sortByCost}\xspace}
\newcommand{\sat}{\mathit{sat}\xspace}
\newcommand{\unique}{\mathit{unique}\xspace}

\newcommand{\isdef}{:=\xspace}
\newcommand{\fsynth}{\phi\xspace}

\newlength{\imgwidth}

\newcommand\scalegraphics[2]{
    \settowidth{\imgwidth}{\includegraphics{#1}}
    \setlength{\imgwidth}{\minof{\imgwidth}{#2}}
    \includegraphics[width=\imgwidth]{#1}
}

\newcommand{\devnull}[1]{}

\newcommand{\VPRE}{\vspace{-0.6em}}
\newcommand{\VPOST}{\vspace{-0.1em}}

\newcommand{\spec}{\ensuremath{\mathit{spec}}\xspace}
\newcommand{\conn}{\ensuremath{\mathit{conn}}\xspace}

\let\origthelstnumber\thelstnumber
\makeatletter
\newcommand*\Suppressnumber{%
  \lst@AddToHook{OnNewLine}{%
    \let\thelstnumber\relax%
     \advance\c@lstnumber-\@ne\relax%
    }%
}

\newcommand*\Reactivatenumber{%
  \lst@AddToHook{OnNewLine}{%
   \let\thelstnumber\origthelstnumber%
   \advance\c@lstnumber\@ne\relax}%
}

\begin{document}

\pagenumbering{arabic}

\title{Efficiently Synthesizing Lowest Cost Rewrite Rules for Instruction Selection }

\author{}

\makeatletter
\newcommand{\linebreakand}{%
  \end{@IEEEauthorhalign}
  \hfill\mbox{}\par
  \mbox{}\hfill\begin{@IEEEauthorhalign}
}

\author{
    \IEEEauthorblockN{Ross Daly}
    \IEEEauthorblockA{
        Stanford University\\
        Stanford, CA, USA\\
        \texttt{rdaly525@cs.stanford.edu}
    }
    \orcid{0000-0002-4938-5250}
    \and
    \IEEEauthorblockN{Caleb Donovick}
    \IEEEauthorblockA{
        Stanford University\\
        Stanford, CA, USA\\
        \texttt{donovick@cs.stanford.edu}
    }
    \orcid{0000-0001-9336-1267}
    \and
    \IEEEauthorblockN{Caleb Terrill}
    \IEEEauthorblockA{
        Stanford University\\
        Stanford, CA, USA\\
        \texttt{cterrill@stanford.edu}
    }
    \and
    \IEEEauthorblockN{Jackson Melchert}
    \IEEEauthorblockA{
        Stanford University\\
        Stanford, CA, USA\\
        \texttt{melchert@stanford.edu}
    }
    \orcid{0000-0002-8232-1603} 
    \and
    \IEEEauthorblockN{Priyanka Raina}
    \IEEEauthorblockA{
        Stanford University\\
        Stanford, CA, USA\\
        \texttt{praina@stanford.edu}
    }
    \orcid{0000-0002-8834-8663}
    \and
    \IEEEauthorblockN{Clark Barrett}
    \IEEEauthorblockA{
        Stanford University\\
        Stanford, CA, USA\\
        \texttt{barrett@cs.stanford.edu}
    }
    \orcid{0000-0002-9522-3084}
    \and
    \IEEEauthorblockN{Pat Hanrahan}
    \IEEEauthorblockA{
        Stanford University\\
        Stanford, CA, USA\\
        \texttt{hanrahan@cs.stanford.edu}
    }
    \orcid{0000-0002-3474-9752}
}

\date{}
\maketitle

\thispagestyle{empty}

\begin{abstract}

Compiling programs to an instruction set architecture (ISA) requires a set of rewrite rules that map patterns consisting of compiler instructions to patterns consisting of ISA instructions.
We synthesize such rules by constructing SMT queries, whose solutions represent two functionally equivalent programs. These two programs are interpreted as an instruction selection rewrite rule. Existing work is limited to single-instruction ISA patterns, whereas our solution does not have that restriction.
Furthermore, we address inefficiencies of existing work by developing two optimized algorithms. The first only generates unique rules by preventing synthesis of duplicate and composite rules. The second only generates lowest-cost rules by preventing synthesis of higher-cost rules.
We evaluate our algorithms on multiple ISAs. Without our optimizations, the vast majority of synthesized rewrite rules are either duplicates, composites, or higher cost. Our optimizations result in synthesis speed-ups of up to 768$\times$ and 4004$\times$ for the two algorithms.

\end{abstract}

\LetLtxMacro{\oldsection}{\section}
\LetLtxMacro{\oldsubsection}{\subsection}
\renewcommand{\section}[2][]{\oldsection{#2}\label{sec:#1}}
\renewcommand{\subsection}[2][]{\oldsubsection{#2}\label{sec:#1}}

\section[intro]{Introduction}

As we approach the end of Moore's law and Dennard scaling, drastically improving computing performance and energy efficiency requires designing domain-specific hardware architectures (DSAs) or adding domain-specific extensions to existing architectures \cite{hennessy}. As a result, many DSAs have been developed in recent years \cite{jouppi2018domain, prabhakar2017plasticine, chen2016eyeriss, bahr2020creating, melchert2023apex}, each with its own custom instruction set architecture (ISA) or ISA extension.
 
Targeting such ISAs from a compiler's intermediate representation (IR) requires a custom library of instruction selection rewrite rules. A rewrite rule is a mapping of an IR pattern to a functionally equivalent ISA pattern. Manual specification of rewrite rules is error-prone, time-consuming, and often incomplete. It is therefore desirable to automatically generate valid rewrite rules.

When specifying instruction selection rewrite rules, there are two common cases. When ISAs have complex instructions, rewrite rules will often map multi-instruction IR patterns to a single ISA instruction. When ISAs have simple instructions, rewrite rules will often map a single IR instruction to a multi-instruction ISA pattern. A rewrite rule generation tool should be able to create rewrite rules for both cases. We call such rewrite rules \textit{\mton} rules.

Generating instruction selectors is not a new idea. Most relevant to this work is Gulwani et al. \cite{gulwani} who use a satisfiability modulo theories (SMT) solver to synthesize a loop-free program that is functionally equivalent to a given specification. 
Their approach is called component-based program synthesis (CBPS), as each synthesized program must include functional components from a given component library. 
Buchwald et al. \cite{buchwald2018synthesizing} use and extend CBPS 
to efficiently generate multi-instruction loop-free IR programs equivalent to a single ISA instruction program; that is, they solve the many-to-one rewrite rules synthesis problem. 
However, multi-instruction ISA programs cannot be synthesized.

Both of these algorithms produce many \textit{duplicate} rules, which are removed during a post-processing step.
As we show, this adds significant additional cost.
Another issue is that CBPS as currently formulated does not incorporate the notion of optimizing for cost. In practice, we often want only the set of lowest-cost rules, making it unnecessary (and expensive) to generate equivalent higher-cost rules. 

This paper presents an algorithm for automatically generating a complete set of \mton rewrite rules. 
We address the above issues by preventing the synthesis of both duplicate and high-cost rules at rule generation time, using
exclusion techniques.
As a further optimization, we generate rules in stages and exclude \emph{composite} rules, i.e. rules that can be composed of smaller rules found in previous stages.
These ensure we produce a minimal but complete set of rewrite rules. Compared to previous work, our approach eliminates unnecessary rules and significantly reduces the time required to produce the unique necessary ones.

Our contributions are as follows:
\begin{itemize}
    \item We define generalized component-based program synthesis (GCBPS) as the task of synthesizing two functionally equivalent programs using two component libraries. We then present an SMT-based synthesis approach inspired by Gulwani et al. to solve it.    
    \item We present an iterative algorithm \genall to generate all unique \mton rules up to a given size. We identify a set of equivalence relations for patterns encoded as programs and for rules that map IR programs to ISA programs. We use these relations to enumerate and exclude duplicate rules. Furthermore, we directly exclude composite rewrite rules. These result in up to a 768$\times$ synthesis speed-up.
    \item We present an algorithm $\genlc$ which generates only the lowest-cost rules by incorporating a cost metric in addition to excluding duplicate and composite rewrite rules.
    This results in a synthesis speed-up up to 4004$\times$.
\end{itemize}

The rest of the paper is organized as follows.
Section~\ref{sec:background} discusses instruction selection, existing rule generation methods, SMT, and program synthesis.
Section~\ref{sec:gcbps} describes a program synthesis query for generating \mton rules. 
Section~\ref{sec:genAll} presents an algorithm for generating only unique rewrite rules and defines duplicates and composites. 
Section~\ref{sec:genallLC} presents an algorithm for synthesizing only the lowest-cost rules.
Section~\ref{sec:eval} evaluates both algorithms, and Section~\ref{sec:conc} discusses limitations and further optimizations.
\section[background]{Background and Related Work}
\label{sec:background}

\subsection{Instruction Selection}
Instruction selection is the task of translating code in the compiler's intermediate representation (IR) to functionally equivalent code for a target ISA.
Typically, a library of rewrite rules is used in instruction selection. 
A rewrite rule is a mapping from an IR pattern consisting of IR instructions to a functionally equivalent ISA pattern consisting of ISA instructions.
Such patterns can be expression trees or directed acyclic graphs (DAGs).

Significant work has been devoted to developing rewrite rule tiling algorithms to perform instruction selection \cite{glan78, gan80, gan82, graham88, emmelmann1989beg, aho1989code, fraser1992engineering, fraser1995retargetable, llvm-codegen, koes2008near}.
For each rule in the rule library, a tiling algorithm first finds all  fragments from the IR program in which the rule's IR pattern exactly matches that fragment.
Then, the instruction selector finds a tiling of these matches that completely covers the basic block and minimizes the total rule cost according to some cost metric.

Simple instruction selectors only handle tree-based IR patterns, which is inefficient for reused computations.
Modern instruction selectors like LLVM, use DAG-based matching that allows for both a richer rules and better tiling. 
Koes et al. \cite{koes2008near} describe a similar near-optimal DAG-based instruction selection algorithm~\cite{llvm-codegen}.
We want to generate rules that can be used with such modern instruction selectors.

\subsection{Generating Instruction Selectors}
Generating instruction selectors from instruction semantics has been a topic of research interest \cite{cattell1980automatic, hoover1996generating,dias2010automatically,daly2022synthesizing,buchwald2018synthesizing}.
Dias and Ramsey \cite{dias2010automatically} introduce an algorithm for generating rewrite rules based on a declarative specification of the ISA.
While this solves part of the \mton rule task, their work relies on an existing set of algebraic rewrite rules for synthesizing semantically equivalent rules.
Our work uses SMT for the instruction and program semantics. However, incorporating certain kinds of algebraic rewrite rules could be an avenue for future optimizations.

Daly et al. \cite{daly2022synthesizing} propose a way to synthesize instruction selection rewrite rules from the register-transfer level (RTL) specification of a processor. 
Their algorithm requires a set of pre-specified IR patterns.
In contrast, we can efficiently synthesize rules that consider all possible multi-instruction IR patterns up to a given size.
Their approach for synthesizing complex instruction constants and handling floating point types could be combined with the approaches in this paper.

The most relevant to this work is the work by Buchwald et al. \cite{buchwald2018synthesizing}, that leverages component-based program synthesis to generate rules with multi-instruction IR patterns and single-instruction ISA patterns. 
In contrast, our work synthesizes rules with both multi-instruction IR patterns and multi-instruction ISA patterns.
We additionally prevent the synthesis of duplicate, composite, and high-cost rewrite rules, unlike any of the above approaches.

\subsection{Program Synthesis and Equivalence}
We use SMT-based program synthesis to enumerate a complete set of instruction selection rewrite rules. In program synthesis enumeration, it is common to remove equivalent solutions \cite{alur2017scaling}. We use the equivalence relation defined in Section~\ref{sec:dup} to determine equivalent rewrite rules. In prior work \cite{albarghouthi2013recursive}, observational equivalence (i.e., programs with the same semantics) has been used for de-duplication
\cite{albarghouthi2013recursive}, however observational equivalence does not take into account the structure of the program which is essential for rewrite rule pattern matching.

\subsection{Logical Setting and Notation}
We work in the context of many-sorted logic (e.g., \cite{enderton2001mathematical}), where we assume an infinite set of variables of each sort. 
Terms are denoted using non-boldface symbols (e.g., $X$). 
Boldface symbols (e.g., $\vv{X}$) are used for sets, tuples, and multisets, whose elements are either terms or other collections of terms.
$\vv{Y} \isdef (Y_1, ..., Y_N)$ defines a tuple, where $|\vv{Y}| = N$ and $Y_i$ refers to the $i$-th element. 
$\vv{Z} \isdef \{z^n\}$ defines a multiset, where the multiplicity of element $z$ is $n \in \mathbb{N}$. Both $\psi$ and $\phi$ are used to denote formulas.
$\psi(\vv{X})$ is a formula whose free variables are a subset of $\vv{X}$.
We use $\model \vDash \psi(\vv{X})$ to denote the \emph{satisfiability} relation between the interpretation $\model$ and the formula $\psi$. Assuming $\vv{X}$ is a collection of variables, $\model_{\vv{X}}$ denotes the \emph{assignment} to those variables induced by $\model$.  For an assignment $\alpha$, we write $\alpha\models\psi(\vv{X})$ if $\model\models\psi(\vv{X})$ for every model $\model$ such that $\model_\vv{X} = \alpha$.

\subsection{Component-based Program Synthesis}

CBPS is a program synthesis task introduced by Gulwani et al. The inputs to the task are:
\begin{itemize}
    \item A \emph{specification} $\vv{S} \isdef (\vv{I}^S, O^S, \phi_{\spec}(\vv{I}^S,O^S))$ containing a tuple of input variables $\vv{I}^S$, a single output variable $O^S$, and a formula $\phi_{\spec}(\vv{I}^S,O^S)$ relating the inputs and the output.
    \item  A \emph{library of components} (e.g., instructions)
    $\vv{K}$, where the $k$-th component $\vv{K}_k \isdef (\vv{I}_k, O_k, \phi_k(\vv{I}_k,O_k))$ consists of a tuple of input variables $\vv{I}_k$, a single output variable $O_k$, and a formula $\phi_{k}(\vv{I}_k,O_k)$ defining the component's semantics. 
\end{itemize}

\newcommand{\bvsort}[1]{\ensuremath{\mathit{BV}_{[#1]}}\xspace}

An example component for an addition instruction is shown below using the theory of bit-vectors, QF\_BV, where \bvsort{n} is an n-bit sort. 
    \[((I_0 : \bvsort{16}, I_1: \bvsort{16}), O : \bvsort{16}, I_0 +_{[16]} I_1 = O) \]

The task is to synthesize a valid program functionally equivalent to the specification using each component from $\vv{K}$ exactly once.

For notational convenience, we group together the set of all inputs and outputs of the components: $\vv{W} \isdef \cup_{(\vv{I}_k, O_k,\_) \in \vv{K}}\left(O_k \cup \left(\cup \vv{I}_k\right)\right)$.
Gulwani et al. encode the program structure using a \emph{connection constraint}: $\phi_{\conn}(\vv{L}, \vv{I}^S, O^S, \vv{W})$. This is a formula representing how the program inputs ($\vv{I}^S$) and program output ($O^S$) are connected via the components.  The connections are specified using \emph{location variables} $\vv{L}$.  We do not go into the details of how location variables encode connections (they are in~\cite{gulwani}).  It is sufficient for our purposes to know that these are integer variables, and an assignment to them uniquely determines a way of connecting the components together into a program.
The \emph{program semantics} $\phi_{\prog}$ are defined as the components' semantics conjoined with the connection constraint:
\begin{align}
    &\phi_{prog}(\vv{L}, \vv{I}^S, O^S, \vv{W}) \isdef \label{eq:prog} \\
    &\ \ \left(\bigwedge_k \phi_k(\vv{I}_k, O_k)\right) \land \phi_{conn}(\vv{L}, \vv{I}^S, O^S, \vv{W}).     
 \nonumber 
\end{align}

They define a verification constraint that holds if a particular program is both well-formed (specified using a well-formedness constraint $\psi_\wfp$) and satisfies the specification $\phi_{\spec}$:
\begin{align}
&\phi_\verif := \psi_\wfp(\vv{L}) \wedge \forall \vv{I}^S, O^S, \vv{W}.  \label{eq:verif1} \\
&\ \ \phi_{prog}(\vv{L}, \vv{I}^S, O^S, \vv{W}) \implies \phi_{spec}(\vv{I}^S, O^S). \nonumber
\end{align}

\noindent
A \emph{synthesis formula} $\phi_{\synth}$ existentially quantifies $\vv{L}$ in~\eqref{eq:verif1}:
\begin{align}
& \phi_{\synth} \isdef \exists \vv{L}. \forall \vv{I}^S, O^S, \vv{W}. \label{eq:cbpssynth} \\
&\ \ \psi_{\wfp}(\vv{L}) \land \left(\phi_{\prog}(\vv{L}, \vv{I}^S, O^S, \vv{W}) \implies \phi_{spec}(\vv{I}^S, O^S)\right). \nonumber
\end{align}

\noindent
This formula can be solved using a technique called counter-example guided inductive synthesis (CEGIS).
CEGIS solves such exist-forall formulas by iteratively solving a series of quantifier-free queries and is often more efficient than trying to solve the quantified query directly. More details are in \cite{gulwani}.  For our purposes, we assume the existence of a CEGIS implementation, $\CEGIS$, which takes an instance of $\phi_\synth$ and returns a model $\model$ with the property that 
$\model_{\vv{L}} \models \phi_\verif$, from which a program that is a solution to CBPS can be constructed.
\section{Component-based Program Synthesis for Many-to-Many Rules}
\label{sec:gcbps}

Given the IR and ISA instruction sets $\ir{K}$ and $\isa{K}$, Buchwald et al. \cite{buchwald2018synthesizing} use \CBPS to synthesize rewrite rules.
They use a single ISA instruction $\isa{k} \in \isa{K}$ for the CBPS specification and a subset of the IR instructions for the CBPS components.
A solution to the resulting $\phi_\synth$ formula gives a program $\ir{P}$.  If $\isa{P}$ is the single-instruction program consisting of $\isa{k}$, they interpret the pair $(\ir{P}, \isa{P})$ as an instruction selection rewrite rule.

However, Buchwald et al.'s solution is insufficient for generating \mton rules, as they cannot synthesize IR and ISA programs that both contain multiple instructions. Instead, two functionally equivalent programs need to be synthesized.
We first define an extension to CBPS called generalized component-based program synthesis (GCBPS) to address this problem. Then we show how to construct a synthesis query whose solutions represent pairs of functionally equivalent programs.

\subsection{Generalized Component-based Program Synthesis}

We define the GCBPS task as that of synthesizing two programs, $\vv{P}^a$ and $\vv{P}^b$, represented using location variables $\vva{L}$ and $\vvb{L}$, given two sets of components $\vva{K}$ and $\vvb{K}$, two sets of inputs $\vva{I}, \vvb{I}$ where $|\vva{I}| = |\vvb{I}|$, and two outputs $O^a, O^b$ where the following conditions hold true:
\begin{enumerate}
    \item $\vva{P}$ uses each component in $\vva{K}$ exactly once.
    \item $\vvb{P}$ uses each component in $\vvb{K}$ exactly once.
    \item $\vva{P}$ is functionally equivalent to $\vvb{P}$.
\end{enumerate}

\subsection{Solving GCBPS}
We start with the CBPS verification constraint from \eqref{eq:verif1} using components $\vva{K}$ (and a corresponding set of inputs and outputs $\vva{W}$), but modify it slightly by introducing variables $(\vva{I}, O^a)$ that are fresh copies of $(\vv{I}^S, O^S)$:

\begin{align}
& \psi_\wfp(\vva{L}) \wedge \forall \vva{I}, O^a, \vva{W}, \vv{I}^S, O^S. \label{eq:verif2} \\
&\ \ (\phi^a_{prog}(\vva{L}, \vva{I}, O^a, \vva{W}) \land \phi_{spec}(\vv{I}^S, O^S)) \implies  \nonumber \\
&\ \ \left(\left(\wedge_{i}\ I^a_i = I^S_i\right) \implies O^a = O^S\right). \nonumber
\end{align}

\noindent
Assuming the formulas for both the program and the specification, if their inputs are the same, their outputs must also be the same.

We next replace the specification program with a different component-based program using components $\vvb{K}$
and quantify over that program's inputs $\vvb{I}$, output $O^b$, and component variables $\vvb{W}$:

\begin{align}
& \phi_\verif := \psi_\wfp(\vva{L}) \wedge \psi_\wfp(\vvb{L}) \wedge \forall \vva{I}, \vvb{I}, O^a, O^b, \vva{W}, \vvb{W}. \label{eq:verif3} \\
&\ \ \left(\phi^a_{prog}(\vva{L}, \vva{I}, O^a, \vva{W}) \land \phi^b_{prog}(\vvb{L}, \vvb{I}, O^b, \vvb{W})\right) \! \! \implies \! \! \nonumber \\
&\ \ \left(\left(\wedge_{i}\ \! I^a_i = I^b_i\right) \! \!  \implies \! \! O^a = O^b\right) \nonumber.
\end{align}

\noindent
This is our generalized verification constraint stating the correctness criteria for when two component-based programs are semantically equivalent.

To synthesize such a pair of programs, a synthesis formula $\phi_{\synth}$ is defined by existentially quantifying $\vva{L}$ and $\vvb{L}$ in the verification formula \eqref{eq:verif3}:
\begin{align}
& \phi_{synth} \isdef \exists \vva{L}, \vvb{L}. \forall \vva{I}, \vvb{I}, O^a, O^b, \vva{W}, \vvb{W}.  \label{eq:gcbps} \\
&\ \ \psi_{wfp}(\vva{L}) \land \psi_{wfp}(\vvb{L}) \land \nonumber \\
&\ \ \biggl( \left(
    \phi^a_{prog}(\vva{L}, \vva{I}, O^a, \vva{W}) \land \phi^b_{prog}(\vvb{L}, \vvb{I}, O^b, \vvb{W}) 
  \right) \implies \nonumber \\
&\ \ \left(
    \left(
        \wedge_{i}\ I^a_i = I^b_i
    \right) \implies O^a = O^b
  \right) \biggr) \nonumber.
\end{align}

\noindent
As above, we assume that calling $\CEGIS$ on $\phi_\synth$ returns a model $\model$ such that $\model_{\vva{L}\cup\vvb{L}}\models \phi_\verif$.  This can be converted into a pair of programs $(\vva{P}, \vvb{P})$ representing a rewrite rule that is a solution for the GCBPS task.  We write $\constructRule(\vva{K}, \vvb{K}, \model_{\vva{L}},\model_{\vvb{L}})$ for the rewrite rule constructed from a specific model $\model$ using the component sets $\vva{K}$ and $\vvb{K}$.

\section{Generating All Many-to-Many Rewrite Rules}
\label{sec:genAll}

Buchwald et al. \cite{buchwald2018synthesizing} describe an iterative algorithm, $\IterativeCEGIS$, to synthesize rewrite rules using CBPS. This algorithm iterates over all multisets of IR instructions up to a given size and only runs synthesis on each such multiset. Compared to running synthesis using all the IR instructions at once, this iterative algorithm works better in practice.

However, $\IterativeCEGIS$ cannot synthesize rewrite rules with both multi-instruction IR programs and multi-instruction ISA programs.
Furthermore, it produces duplicate rewrite rules which are then filtered out in a post-synthesis filtering step. Although the results are correct, this approach is highly inefficient because each call to CEGIS is expensive, and a CEGIS call is made, not just for some duplicate rules, but for every possible duplicate rule.  In our approach, we make the requirement that a solution is not a duplicate part of the CEGIS query itself, ensuring that each successful CEGIS query finds a new, non-redundant rewrite rule.

\begin{figure}[t]
    \centering

    \begin{minipage}{\linewidth}
\begin{lstlisting}[mathescape=true, xleftmargin=2em]
$\genall(\ir{K}, \isa{K}, N^\mathit{IR}, N^\mathit{ISA})$:
    $\SR \leftarrow \{\} $
    for $n_1, n_2 \in [1,N^{\mathit{IR}}] \times [1,N^{\mathit{ISA}}]$:
        for $\ir{m} \in \multicomb(\ir{K},n_1)$:
            for $\isa{m} \in \multicomb(\isa{K},n_2)$:
                for $\ir{I}, \isa{I} \in \allInputs(\ir{m}, \isa{m})$:
                    $\fsynth, \ir{L}, \isa{L} \leftarrow$ |\Suppressnumber|
                        $\GCBPS(\ir{m}, \isa{m}, \ir{I}, \isa{I})$ |\Reactivatenumber|
                    $\fsynth \leftarrow \fsynth\ \wedge \neg \AllC(\SR, \ldots)$
                    $\SR\ \leftarrow \SR\ \cup$ |\Suppressnumber|
                        $\CEGISAll(\fsynth, \ir{m}, \isa{m}, \ir{L}, \isa{L})$ |\Reactivatenumber|
    return $\SR$
\end{lstlisting}
\caption{Iterative algorithm to generate all unique rewrite rules up to a given size.}
\label{alg:genAll}
\end{minipage}
\VPOST


\begin{minipage}{\linewidth}
\begin{lstlisting}[mathescape=true,  xleftmargin=2em]
$\CEGISAll(\fsynth, \ir{m}, \isa{m}, \ir{L}, \isa{L})$:
    $\SR = \{\}$
    while True:
        $\model \leftarrow \CEGIS(\fsynth)$
        if $\model = \bot$: return $\SR$
        $\vv{R} \leftarrow \constructRule(\ir{m},\isa{m},\model_{\ir{L}}, \model_{\isa{L}})$
        $\SR \leftarrow \SR \cup \{\vv{R}\}$
        $\fsynth \leftarrow \fsynth \land \neg \RuleL(\vv{R}, (\ir{L}, \isa{L}))$

\end{lstlisting}
\caption{AllSAT algorithm to synthesize all unique rules. Line 8 excludes all rules that are duplicates of the current synthesized rewrite rule.}
\label{alg:cegisAll}
\end{minipage}
\end{figure}

Our iterative algorithm, \genall, is shown in Figure~\ref{alg:genAll}.  It takes as parameters the IR and ISA component sets, $\ir{K}$ and $\isa{K}$ respectively, as well as a maximum number of components of each kind to use in rewrite rules, $N^\mathit{IR}$ and $N^\mathit{ISA}$, and iteratively builds up a set $\SR$ of rewrite rules, which it returns at the end.
Line 3 shows that $n_1$ and $n_2$ iterate up to these maximum sizes.  Line 4 iterates over all multisets of elements from $\ir{K}$ of size $n_1$ using a standard multicombination algorithm $\multicomb$~\cite{knuth} (not shown). Line 5 is similar but for multisets from $\isa{K}$ of size $n_2$.
Next, for a given choice of multisets, line 6 enumerates all possible ways of selecting input vectors from those multisets that could create well-formed programs. Line 7 constructs fresh sets of location variables $\ir{L}$ and $\isa{L}$ and returns them along with the instantiated GCBPS synthesis formula (using Equation~\eqref{eq:gcbps}).\footnote{We augment the well-formed program constraint in \eqref{eq:gcbps} to prevent synthesizing programs containing dead code and unused inputs. This can be accomplished by enforcing that each input and intermediate value is used in at least one location.}
Line 8 excludes all \textit{composite rules} from the synthesis search space. Composite rules are rules that can be constructed using the current set of rules $\SR$ and are thus unnecessary for instruction selection. We discuss this in more detail in Section \ref{sec:comp}.
Finally, on line 9, the current set of rules $\SR$ is updated with the result of calling $\CEGISAll$, which we describe next.

Figure~\ref{alg:cegisAll} shows the $\CEGISAll$ algorithm that performs the AllSAT \cite{allsat, allsat2} task. Its parameters are the synthesis formula $\phi$, the multisets $\ir{m}$ and $\isa{m}$, and the location variables $\ir{L}$ and $\isa{L}$.  It returns a set $\SR$ of rewrite rules.  Initially this set is empty.  The algorithm iteratively calls a standard $\CEGIS$ algorithm to solve the synthesis query, constructing a new rewrite rule $\vv{R}$, which is added to the set $\SR$ of rewrite rules, when the call to $\CEGIS$ is successful. The iteration repeats until the $\CEGIS$ query returns $\bot$, indicating that there are no more rewrite rules to be found.  Note that after each iteration, the $\phi_\synth$ formula is refined by adding the negation of a formula capturing the notion of duplicates for this rule.  We describe how this is done next.

\subsection{Excluding Duplicate Rules}
\label{sec:dup}
Consider the two distinct rules below. As a syntactical convention, infix operators are used for IR patterns and function calls for ISA patterns.
\begin{align*}
    I_1 + (I_2 \cdot I_3) \to add(I_1, mul(I_2, I_3))\\
    (I_1 \cdot I_3) + I_2 \to add(I_2, mul(I_1, I_3))
\end{align*}
The two IR patterns represent the same operation despite the fact that the variable names and the order of the commutative arguments to addition are both different.
Both rules would match the same program fragments in an instruction selector and would result in the same rewrite rule application. Thus, we consider such rules to be equivalent and would like to ensure that only one is generated by our algorithm.

We first define a rewrite rule equivalence relation, $\ruleeq$. Informally, two rules are equivalent if replacing either one by the other has no discernible effect on the execution of an instruction selection algorithm. We make this more formal by considering various attributes of standard instruction selection algorithms. 

\newcommand{\simC}[1]{\sim_{C^{\mathit{#1}}}\xspace}
\newcommand{\simK}[1]{\sim_{K^{\mathit{#1}}}\xspace}
\newcommand{\simD}[1]{\sim_{D^{\mathit{#1}}}\xspace}
\newcommand{\simI}[1]{\sim_{I^{\mathit{#1}}}\xspace}

\smallskip
\noindent
\textbf{Commutative Instructions}
Modern pattern matching algorithms used for instruction selection try all argument orderings for commutative instructions~\cite{llvm-codegen}. We define the commutative equivalence relation $\simC{IR}$ as $\ir{P}_1 \simC{IR} \ir{P}_2$ iff $\ir{P}_2$ is a remapping of $\ir{P}_1$'s commutative instruction's arguments. 

\smallskip
\noindent
\textbf{Same-kind Instructions}
Programs $\vv{P}$ generated by GCPBS have a unique identifier, the program line number, for each instruction. This means that if two instructions of the same kind appear in a program, interchanging their line numbers results in a different program, even though it makes no difference to the instruction selection algorithm.
We define the same-kind equivalence relation $\simK{IR}$ as $\ir{P}_1 \simK{IR} \ir{P}_2$ iff $\ir{P}_2$ is the result of remapping the line numbers for same-kind instructions in $\ir{P}_1$.

\smallskip
\noindent
\textbf{Data Dependency}
Modern instruction selection algorithms perform pattern matching, not based on a total order of instructions, but on a partial order determined by data dependencies.  Many different sequences may thus lead to the same partial order.  We define $\simD{IR}$ as $\ir{P}_1 \simD{IR} \ir{P}_2$ iff $\ir{P}_1$ and $\ir{P}_2$ have the same data dependency graph.

\smallskip
\noindent
\textbf{Rule Input Renaming}
For a given rewrite rule, the input variables used for the IR program must match the input variables used for the ISA program, but the specific variable identifiers used do not matter.  We define the equivalence relation $\simI{rule}$ on rules (i.e., pairs of programs) as $\vv{R}_1 \simI{rule} \vv{R}_2$ iff $\vv{R}_2$ is the result of remapping variable identifiers in $\vv{R}_1$.

\smallskip
\noindent
\textbf{Rule Equivalence}
The first three equivalence relations defined above are for IR programs, but the analogous relations ($\simC{ISA}$, $\simK{ISA}$, $\simD{ISA}$) for ISA instructions are also useful.

Putting everything together, we define rule equivalence $\ruleeq$ as follows.
\begin{align}
    \sim_{\mathit{IR}}\ &\isdef  \uplus\{\simC{IR}, \simK{IR}, \simD{IR}\} \label{eq:ireq1}\\
    \sim_{\mathit{ISA}}\ &\isdef \uplus\{\simC{ISA}, \simK{ISA}, \simD{ISA}\} \\
    \ruleeq\ &\isdef \uplus\{(\sim_{\mathit{IR}} \otimes \sim_{\mathit{ISA}}), \simI{rule}\} \label{eq:ruleeq}
\end{align}

\noindent
Overall IR equivalence is defined as the transitive closure of the union (notated with $\uplus$) of the three individual IR relations. ISA equivalence is defined similarly.  Overall rewrite rule equivalence is then defined using the $\otimes$ operator, where $\sim_\otimes = \sim_a \otimes \sim_b$ is defined as: $(a_1,b_2) \sim_\otimes (a_2,b_2)$ iff $a_1 \sim_a a_2$ and $b_1 \sim_b b_2$.  Specifically, rule equivalence is obtained by combining IR equivalence in this way with ISA equivalence, and then combining the result with $\simI{rule}$ using $\uplus$.

\newcommand{\simDCE}[1]{\sim_{DCE^{#1}}\xspace}
\newcommand{\simCSE}[1]{\sim_{CSE^{#1}}\xspace}

The set of all duplicates of rule $\vv{R}$ is the rule equivalence class $\ruleset{\vv{R}}$, where $\vv{R'} \in \ruleset{\vv{R}} \iff \vv{R} \ruleeq \vv{R'}$. $\RuleL$ can be constructed by enumerating all elements of the equivalence class $\ruleset{\vv{R}}$.

\subsection{Excluding Composite Rules}
\label{sec:comp}
We also exclude any rule whose effect can already be achieved using the current set of generated rules (line 8 of Figure~\ref{alg:genAll}).
We elucidate this using a simple example. 
Assume the algorithm just constructed a new query for the multisets $\ir{m}$, $\isa{m}$, and the input $\ir{I}$ (line 7 of Figure~\ref{alg:genAll}) and assume that
the rule library $\SR$ currently contains rules for addition (\mbox{$I_1 + I_2 \to add(I_1, I_2)$}), and multiplication (\mbox{$I_1 \cdot I_2 \to mul(I_1, I_2)$}). Consider the following cases.

\begin{enumerate}
    \item If $\ir{I} = (I_1)$, $\ir{m} = \{+\}$, and  $\isa{m} = \{add\}$, then the rule $I_1 + I_1 \to add(I_1, I_1)$ will be synthesized by $\CEGISAll$. But this rule is a \textit{specialization} of the existing rule for addition. Any use of this specialized rule could instead be replaced by the more general rule and this rule can thus be excluded.
    Note that we order the inputs on line 6 of Figure~\ref{alg:genAll} to guarantee that the most general version of a rule is found first. \label{case:all1}
    \item If $\ir{I} = (I_1, I_2, I_3)$, $\ir{m} = \{+, \cdot\}$, and  $\isa{m} = \{add, mul\}$, then the composite rule $(I_1 + (I_2 \cdot I_3)) \to add(I_1, mul(I_2, I_3))$ will be synthesized by $\CEGISAll$. Using similar logic, any use of this composite rule could instead use the simpler and more general rules for addition and multiplication, and this rule can thus be excluded. The multiset ordering used in lines 4 and 5 of Figure~\ref{alg:genAll} ensures that subsets are visited before supersets, guaranteeing that smaller rules are found first. 
\end{enumerate}

\smallskip
\noindent
Only a subset of composite rules built from existing rules need to be excluded for each synthesis query.  In general, for a specific query based on $\ir{m}$, $\isa{m}$, and $\ir{I}$, we exclude composite rules $\vv{R} \isdef (\ir{P}, \isa{P})$ that meet the following criteria:
\begin{itemize}
    \item $\vv{R}$ has exactly $|\ir{I}|$ inputs.
    \item $\ir{P}$ has the same components as $\ir{m}$.
    \item $\isa{P}$ has the same components as $\isa{m}$.
    \item $\ir{P}$ is built from the IR programs of already-found rules in $\SR$.
    \item $\isa{P}$ is the result of applying the rewrite rules used to build $\ir{P}$.
\end{itemize}
These checks are encapsulated by the call to $\AllC$ on line 8 of Figure~\ref{alg:genAll}.

\section{Generating All Lowest-Cost Rules}
\label{sec:genallLC}

\begin{figure}[t]
\begin{lstlisting}[mathescape=true,  xleftmargin=2em]
$\genall_{LC}(\ir{K}, \isa{K}, N^\mathit{IR}, N^\mathit{ISA}, \cost)$:
    $\vv{K}_{\sorted} \leftarrow \sortByCost(\isa{K}, N^{ISA}, \cost)$
    $\SR \leftarrow \{\}$
    for n $\in [1,N^\mathit{IR}]$:
        for $\ir{m} \in \multicomb(\ir{K},n)$:
            for $\isa{m} \in \vv{K}_{\sorted}$:
                $c_{cur} \leftarrow \cost(\isa{m})$
                for $\ir{I}, \isa{I} \in \allInputs(\ir{m}, \isa{m})$:
                    $\fsynth, \ir{L}, \isa{L} \leftarrow$ |\Suppressnumber|
                        $\GCBPS(\ir{m}, \isa{m}, \ir{I}, \isa{I})$ |\Reactivatenumber|
                    $\fsynth \leftarrow \fsynth \wedge \neg \AllC_\mathit{LC}(\SR, c_{cur}, \ldots)$
                    $\SR\ \leftarrow \SR\ \cup$ |\Suppressnumber|
                        $\CEGISAll_\mathit{LC}(\fsynth, \ir{m}, \isa{m}, \ir{L}, \isa{L})$ |\Reactivatenumber|
    return $\SR$
\end{lstlisting}


\VPRE
\caption{Iterative algorithm to generate all lowest-cost rules. ISA multisets are ordered by cost. $\CEGISAll$ is modified to exclude rules with duplicate IR programs.}
\label{alg:genAll2}
\VPOST
\end{figure}

Because all duplicates are excluded, the $genAll$ algorithm generates only unique rewrite rules.
However, two unique rules can share the same IR pattern. For a particular IR pattern, only the lowest-cost rule is needed for some cost metrics.
Knowing the instruction selection cost metric at rule-generation time presents another time-saving opportunity because we can also prevent the synthesis of high-cost rules.

We make a few assumptions about such a cost metric. 
\begin{itemize}
    \item The cost for an instruction selection tiling is equal to the sum of the costs of each tiling rule's ISA program.
    \item The cost of an ISA program $\isa{P}$ only depends on the instruction \textit{contents}, not the program structure. This cost is the sum of the cost of each instruction in the program.
\end{itemize}
While these assumptions are a restriction on the space of possible cost metrics, they are sufficient to represent common ones like code size and energy. If the compiler's cost metric violates these assumptions, the \genall algorithm can be used instead. 
This restricted space of cost metrics has the important property that the cost of any rule that would be synthesized using the components $\isa{m}$ can be determined up front as the sum of the cost of each component.

Figure~\ref{alg:genAll2} shows our synthesis algorithm updated to only synthesize the lowest-cost rules for each unique IR pattern.
The first change is to sort all possible mulitsets of ISA instructions up to size $N^\mathit{ISA}$ by cost (lower cost first) (line 2).
This ordering ensures that the first rule synthesized for a particular IR program will be the lowest-cost version of that rule.
Therefore, after synthesizing a new rule, all rules with a duplicate IR program can be excluded. 
The second change excludes rules with duplicate IR programs.
A duplicate IR program is defined using the IR equivalence relation:
\begin{align}
    \sim_{\mathit{IR}_\mathit{LC}} &\isdef \uplus\{\simC{IR}, \simK{IR}, \simD{IR}, \simI{IR}\} \label{eq:eqirlc}
\end{align}
This is the same definition as \eqref{eq:ireq1}, but with an additional relation $\simI{IR}$ defined as $\ir{P}_1 \simI{IR} \ir{P}_2$ iff $\ir{P}_2$ is the result of remapping variable identifiers in $\ir{P}_1$.  The $\CEGISAll_\mathit{LC}$ function called on line 11 is the same as $\CEGISAll$, except that it uses $\sim_{\mathit{IR}_\mathit{LC}}$ instead of $\sim_\mathit{IR}$ when constructing $\RuleL$.

The third change modifies $\AllC$ to use the known up-front cost $\cost(\isa{m})$.
To see how this works, we consider again the example from Section~\ref{sec:comp}. As before, we assume
$\SR$ currently contains two rules: one for addition (\mbox{$I_1 + I_2 \to add(I_1, I_2)$}) and one for multiplication (\mbox{$I_1 \cdot I_2 \to mul(I_1, I_2)$}).
We assume the target (ISA) expressions for these rules have cost 5 and 10, respectively.  Consider the following situation:

\begin{itemize}
    \item Suppose $\ir{I} = (I_1, I_2, I_3)$, and $\ir{m} = \{+, \cdot\}$. It might be possible to synthesize a rule that has IR pattern $(I_1 + (I_2 \cdot I_3))$. We know that the composite rule $(I_1 + (I_2 \cdot I_3)) \to add(I_1, mul(I_2, I_3))$ would have a cost of 15 since rule costs are additive. Therefore, we can exclude any rule that matches this IR pattern and has $\cost(\isa{m}) \ge 15$.
\end{itemize}
To implement this, only one adjustment needs to be made to the conditions in Section~\ref{sec:comp}. Instead of requiring $\isa{P}$ to have the same components as $\isa{m}$, we simply require $\cost(\isa{P}) \geq \cost(\isa{m})$, i.e., for rules matching the other conditions, if the ISA program has a cost equal to or greater than cost of the ISA program in the current rule, it is excluded.
These conditions are encapsulated by the call to $\AllC_\mathit{LC}$ (line 10).
\section[eval]{Evaluation}
\label{sec:eval}
Our evaluation strategy is threefold. We first show that our algorithm is capable of producing a variety of \mton rules.  
A good set of rewrite rules involves both many-to-one and one-to-many rules. 
We also show that by removing duplicate, composite, and high-cost rules, we produce a much smaller set of rewrite rules. 
Second, we analyze the effect on performance of the optimizations described above. We show that they all significantly reduce the time spent in synthesis.  
Finally, we show that by using different cost metrics, we can generate different sets of lowest-cost rewrite rules.

\subsection{Implementation}

All instructions are formally specified using the hwtypes Python library \cite{hwtypes}, which leverages pySMT \cite{pysmt} to construct (quantifier-free) SMT queries in the theory of bit-vectors.
We also use annotations indicating which instructions are commutative.
We use Boolector \cite{btor} as the SMT solver and set a timeout of 12 seconds for each CEGIS invocation.
Every synthesized rewrite rule is independently verified to be valid. %

\subsection{Instruction Specifications}
To evaluate our algorithms, we selected small but non-trivial sets of IR and ISA instructions operating on 4-bit bit-vectors.

\smallskip
\noindent
\textbf{IR}
We define the IR instruction set to be constants (0, 1), bitwise operations ($not$, $and$, $or$, $xor$), arithmetic operations ($neg$, $add$, $sub$), multiplication ($mul$), unsigned comparison operations ($ult$, $ule$, $ugt$, $uge$), equality ($eq$), and dis-equality ($neq$). 

\smallskip
\noindent
\textbf{ISA 1}
This is a minimal RISC-like ISA containing only 6 instructions: $nand$, $sub$, three comparison instructions ($cmpZ$, $cmpN$, $cmpC$) which compute the zero (Z), sign (N), and carry (C) flags respectively for a subtraction, and a flag inverting instruction ($inv$). 

\smallskip
\noindent
\textbf{ISA 2}
This is an ISA specialized for linear algebra. It supports the 5 instructions: $neg$, $add$, $add3$ (addition of 3 values), $mul$, and $mac$ (multiply-accumulate).

\newcommand{\genA}{All Rules\xspace}
\newcommand{\genB}{Only Unique\xspace}
\newcommand{\genC}{Only Lowest-Cost\xspace}

    
\begin{table*}[t]
    \centering
    \begin{tabular}{|c c|c c c||c c c||c c c c c|}
    \cline{3-13}
    \multicolumn{2}{c|}{} &\multicolumn{11}{c|}{ISA Program Size} \\
    \cline{3-13}
    \multicolumn{2}{c|}{} & \multicolumn{3}{c||}{\genA} & \multicolumn{3}{c||}{\genB} & \multicolumn{5}{c|}{\genC} \\
    \multicolumn{2}{c|}{} & 1 & 2 & 3 &  
        1 & 2 & 3 & 
        1 & 2 & 3 & 4 & 5 \\ 
    \hline
    \multirow{2}{40pt}{IR Prog Size} 
    & 1 & 
        5 & 32 & 1096 & 
        3 & 10 & 96 &      
        3 & 4 & 2 & 1 & 0 \\ 
    & 2 &
        76 & 1719 & 56894 & 
        40 & 189 & 1940 & 
        40 & 67 & 34 & 12 & 6 \\ 
    \hline
    \end{tabular}
    \caption{Number of synthesized rewrite rules for ISA 1a.}
    \label{tab:nm-ab}

    \begin{tabular}{|c c|c c c||c c c||c c c c|}
    \cline{3-12}
    \multicolumn{2}{c|}{} &\multicolumn{10}{c|}{ISA Program Size} \\
    \cline{3-12}
    \multicolumn{2}{c|}{} & \multicolumn{3}{c||}{\genA} & \multicolumn{3}{c||}{\genB} & \multicolumn{4}{c|}{\genC} \\
    \multicolumn{2}{c|}{} & 1 & 2 & 3 & 
        1 & 2 & 3 &
        1 & 2 & 3 & 4 \\ 
    \hline
    \multirow{2}{40pt}{IR Program Size} 
    & 1 & 
        17 & 71 & 3662 & 
        9 & 51 & 873 &  
        7 & 3 & 0 & 0 \\ 
    & 2 &
        89 & 3942 (-5) & 199572 & 
        78 & 717 & 21511 &  
        52 & 64 & 9 & 0 \\ 
    \hline
    \end{tabular}
    \caption{Number of synthesized rewrite rules for ISA 1b.}
    \label{tab:nm-cmp}
    
    \begin{tabular}{|c c|c c c||c c c||c c c|}
    \cline{3-11}
    \multicolumn{2}{c|}{} & \multicolumn{9}{c|}{IR Program Size} \\
    \cline{3-11}
    \multicolumn{2}{c|}{} & \multicolumn{3}{c||}{\genA} & \multicolumn{3}{c||}{\genB} & \multicolumn{3}{c|}{\genC} \\
    \multicolumn{2}{c|}{} & 
        1 & 2 & 3 &  
        1 & 2 & 3 &
        1 & 2 & 3 \\ 
    \hline
    \multirow{2}{50pt}{ISA Program Size} 
    & 1 & 
        11 & 287 & 3998 & 
        3 & 14 & 315 & 
        3 & 14 & 315 \\ 
    & 2 &
        10 & 3115 & $341758^*$ &  
        3 & 69 & 1337 & 
        1 & 32 & 760 \\ 
    \hline
    \end{tabular}
    \caption{Number of synthesized rewrite rules for ISA 2.}
    \label{tab:nm-cisc}
\end{table*}

\begin{table}[t]
\centering
\begin{tabular}{|c c|c c|}
    \hline
    ISA & Rule Size up & \% Duplicate & \% High-cost \\
        & to (IR, ISA) & or Composite &  \\
    \hline
    1a & (2, 3) & 96.2\% & 99.7\% \\ 
    1b & (2, 3) & 88.8\% & 99.9\% \\ 
    2  & (3, 2) & 99.5\% & 99.7\% \\
    \hline
    \end{tabular}
    \caption{Percent of rewrite rules up to (IR, ISA) size that are a duplicate or a composite, and percent that are high-cost.}
    \label{tab:percentage}
\end{table}

\subsection{Rewrite Rule Synthesis}
For each ISA we run three experiments.
The first experiment (\genA) is the baseline that generates all many-to-many rules including duplicate, composite, and high cost rules. This is an implementation of Buchwald et al.'s \IterativeCEGIS algorithm extended to use GCBPS for many-to-many rules (notated as \baseline).
The second (\genB) generates only unique rules by excluding all duplicates and composites using the $\genall$ algorithm. 
The third (\genC) generates only the lowest-cost rules using the $\genlc$ algorithm in Figure~\ref{alg:genAll2}. A code-size cost metric is used, i.e., $\cost(\vv{K})$ is just the number of components in $\vv{K}$.

For ISA 1, we split the rule generation into two parts. The first part (ISA 1a) synthesizes rules composed of bitwise and arithmetic IR instructions using the ISA's $nand$ and $sub$ instructions. The second part (ISA 1b) synthesizes rules composed of constants and comparison instructions using the four instructions $cmpZ$, $cmpN$, $cmpC$, and $inv$.

For 1a and 1b, we synthesize rewrite rules up to an IR program size of 2 and an ISA program size of 3 (written 2-to-3). For (\genC), we increase the ISA program size to 5 and 4 respectively. For ISA 2, we synthesize all rewrite rules composed of constant, and arithmetic (including $mul$) IR instructions up to size 3-to-2.

The number of rewrite rules produced for ISA 1a, 1b, and 2 are shown in Tables~\ref{tab:nm-ab},~\ref{tab:nm-cmp}, and~\ref{tab:nm-cisc}, respectively.
Each table entry is the number of rewrite rules synthesized for a particular IR and ISA program size. 
For all ISAs, the extra synthesized rules in (\genA) were compared against the duplicate and composite rules excluded by (\genB).
Entries in (\genA) marked with a `(-n)' represent `n' rules that (\genB) synthesized, but (\genA) missed due to CEGIS timeouts. The (\genA) experiment for the entry marked with an asterisk could not complete in 70 hours, so the number calculated from (\genB) is shown.


For both ISAs we were able to synthesize 1-to-many and many-to-1 rules for both IR and ISA instructions. \genall produced a more complete set of rules than \baseline.

Table~\ref{tab:percentage} shows the percentage of rules that are duplicates or composites in the first column, and the percentage of rules that are high cost in the second column.
Most rules in (\genA) are duplicates, composites, or high cost. 
Out of the 349179 rules up to size 3-to-2 for ISA 2 (i.e. the sum of the (\genA)), 99.5\% are duplicates or composites.
Similarly, most rules are high cost. In ISA 1a, 59672 out of 59822 rules (99.7\%) up to size 2-to-3 are high cost. 

\subsection{Synthesis Time Improvement with \genall}
In this section we showcase the synthesis time improvements of \genall. The first experiment is the baseline \baseline. The second excludes duplicate rules (i.e., with line 8 of Figure~\ref{alg:cegisAll}). The third, \genall, excludes both duplicates and composites (i.e. with line 8 of both Figure~\ref{alg:cegisAll} and Figure~\ref{alg:genAll}).

For each $\GCBPS$ query, we note the time required ($t_{\sat}$) to run $\CEGISAll$.
Next, we measure the number of unique rules ($N_{\unique}$) found by $\CEGISAll$.
We then add the pair ($N_{\unique}, t_{\sat}$) to our dataset.
We plot the cumulative synthesis time versus the number of unique rules found by doing the following. 
Each data point is sorted by its slope ($t_{\sat}/N_{\unique}$). Then, the increase in both $t_{\sat}$ and $N_{\unique}$ is plotted for each sorted point. 
Some data points have $N_{\unique} = 0$ indicating that every synthesized rule was redundant and is shown using a vertical slope.

The synthesis time plot for unique rewrite rules for ISA 1b up to size 2-to-3 is shown in Figure \ref{fig:dup}. 
Excluding all duplicates shows a 5.3$\times$ speedup. 
Excluding both duplicates and composites shows a 6.2$\times$ speedup. Both optimizations find an additional 5 unique rules.

\begin{figure*}[htbp]
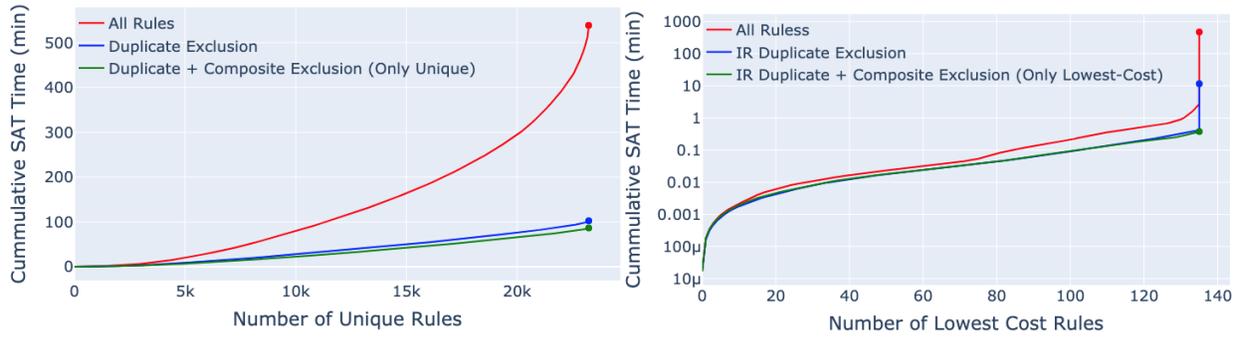

    \centering
    \begin{subfigure}{.45\linewidth}
        \centering
        \scalegraphics{{figs/cmp-uniq-fig}.png}{\linewidth}
        \caption{$\genall$}
        \label{fig:dup}
    \end{subfigure}%
    \begin{subfigure}{.45\linewidth}
        \centering
	\scalegraphics{{figs/cmp-lc-fig}.png}{\linewidth}
	\caption{$\genlc$}
	\label{fig:lc}
    \end{subfigure}
    \caption{Cumulative synthesis time comparison for ISA 1b up to size 2-to-3.}
\end{figure*}

\subsection{Synthesis Time Improvement with \genlc}
We also showcase the synthesis time improvements of \genlc using a similar setup.
The first experiment is the baseline \baseline. The second excludes IR duplicate rules. The third, \genlc, excludes both IR duplicates and IR composites.

We use the same experimental setup as before except when computing $N_{\unique}$, all higher-cost rules are filtered instead.
The synthesis time plot for lowest-cost rewrite rules for ISA 1b up to size 2-to-3 is shown in Figure \ref{fig:lc}.

Excluding rules with duplicate IR programs provides a 41$\times$ speed-up.
Also excluding high-cost composites provides a 1254$\times$ speed-up over the baseline (\genA) configuration. 

\subsection{Total Speed-up}
We summarize the speed-ups of $\genall$ and $\genlc$ compared to the \baseline baseline for all configurations in Table~\ref{tab:speed}. 
We compare the synthesis time in the ``Synth'' column. We compare the total algorithm runtime in the ``Total'' column (including time for iterating, solving, rule filtering, etc.). 
The last row's baseline did not complete in 70 hours, so we provide lower bounds for speed-up.

The speed-ups depend on many parameters including the maximum size of the rewrite rules, the number of possible instructions, the commutativity of the instructions, and the semantics of the instructions.
The optimizations discussed produce several orders of magnitude speed-ups. Further optimizing the non-solver portions (e.g. re-coding in C) would drastically increase the    ``Total'' speed-ups to be closer to the ``Synth'' ones.
Clearly, the combination of all optimizations discussed in this paper can produce speed-ups of several orders of magnitude.

\begin{table}
    \begin{center}
    \begin{tabular}{|c c|r r||r r|}
    \hline
    ISA & Rule Size up & \multicolumn{2}{c||}{$\genall$ Speed-up} & \multicolumn{2}{c|}{$\genlc$ Speed-up} \\
    \cline{3-6}
        & to (IR, ISA) & Synth & Total & Synth & Total \\
    \hline
    1a & (2, 2) & 3.5$\times$ & 1.3$\times$ & 11$\times$ & 2.8$\times$ \\
    1b & (2, 2) & 3.1$\times$ & 1.7$\times$ & 26$\times$ & 2.8$\times$  \\
    2 & (2, 2) & 11$\times$ & 2$\times$ & 53$\times$ & 2.5$\times$ \\
    1a & (2, 3) & 12$\times$ & 6.8$\times$ & 601$\times$ & 57$\times$ \\
    1b & (2, 3) & 6.2$\times$ & 2.7$\times$ & 1254$\times$ & 63$\times$ \\
    2  & (3, 2) & \textgreater{} 768$\times$ & \textgreater{} 81$\times$ & \textgreater{} 4004$\times$ &\textgreater{} 171$\times$ \\
    \hline
    \end{tabular}
    \caption{Speed-ups compared to \baseline.}
    \label{tab:speed}
    \end{center}
\end{table}

\subsection{Cost Metric Comparisons}
Our final experiment explores how the choice of cost metric influences the rules.
We have implemented two cost metrics: a code size metric (CS) and an estimated energy metric (E). 
The energy metric was created to correspond to real hardware energy data. 
For example the cost ratio for $mul$ and $add$ is $1:1$ for code size, but is $2.5:1$ for energy. 
The number of common and unique lowest-cost rewrite rules for each ISA is shown in Table~\ref{tab:cost}.

\begin{table}
    \begin{center}
    \begin{tabular}{|c c|c c c|}
    \hline
    ISA & Rule Size up & Unique & Unique & Common \\
    & to (IR, ISA) & (CS) & (E) &  \\
    \hline
    1a & (2, 5) & 121 & 161 & 48 \\
    1b & (2, 4) & 99 & 198 & 36 \\
    2  & (3, 2) & 134 & 137 & 991 \\
    \hline
    \end{tabular}
    \end{center}
    \caption{Number of unique and common rewrite rules synthesized for code size (CS) and energy (E) cost metrics.}
    \label{tab:cost}
\end{table}

While there is some overlap in common rules, each cost metric produces a differing set of unique lowest-cost rules.

\section[conclusion]{Conclusion and Future Work}
\label{sec:conc} 
We showed that \mton instruction selection rewrite rules can be synthesized for various ISAs using program synthesis.
This supports two major trends in computer architecture. The first is the trend towards simple or reduced instruction architectures where multiple instructions are needed for simple operations. 
It also supports the trend to introduce more complex domain-specific instructions for energy efficiency. In this case, a single instruction can implement complex operations.

We showed that our algorithms are efficient. Removing duplicates, composites, and higher-cost rules results in multiple orders of magnitude speed-ups.
Synthesizing \mton rewrite rules for modern IRs and ISAs may require further optimizations. Many of our synthesized rules contain program fragments that a compiler would optimize before instruction selection (e.g., $sub(X,X)$). Excluding these could result in further speed-ups.

Buchwald et al. \cite{buchwald2018synthesizing} presented generalizations for multi-sorted instructions, multiple outputs, preconditions, and internal attributes, enabling the modeling of memory and control flow instructions. 
Our synthesis query and algorithms are orthogonal and could incorporate these features, allowing for a broader range of possible instruction sets. 

As is the case in prior work, we limit synthesis to loop free patterns. Relaxing this constraint and using other instruction selection algorithms would be an interesting research avenue.

We believe this research area is fertile ground and hope our work inspires and enables future research endeavors towards the goal of automatically generating compilers for emerging domain-specific architectures.

\LetLtxMacro{\section}{\oldsection}
\LetLtxMacro{\subsection}{\oldsubsection}

\bibliographystyle{plain}
\bibliography{references}
\newpage
\clearpage

\end{document}